\documentclass[aps,prl,twocolumn,preprintnumbers,amsmath,amssymb,superscriptaddress,10pt]{revtex4-2}
\usepackage{amsmath,graphicx}
\usepackage[utf8]{inputenc}
\usepackage[T1]{fontenc}
\usepackage{xcolor}
\usepackage{textcomp}
\usepackage{bm}

\usepackage{physics}
\usepackage{amsmath}
\usepackage{tikz}
\usepackage{mathdots}
\usepackage{yhmath}
\usepackage{cancel}
\usepackage{color}
\usepackage{array}
\usepackage{multirow}
\usepackage{amssymb}
\usepackage{gensymb}
\usepackage{tabularx}
\usepackage{extarrows}
\usepackage{booktabs}
\usetikzlibrary{fadings}
\usetikzlibrary{patterns}
\usetikzlibrary{shadows.blur}
\usetikzlibrary{shapes}

\usepackage{color}
\definecolor{LinkColor}{rgb}{0.75,0.0,0.2}

\usepackage{hyperref}
\hypersetup{
	pdfauthor={good guys},
	pdftitle={good title},
	colorlinks=true,
	citecolor=LinkColor,
	linkcolor=LinkColor,
	urlcolor=LinkColor,
}

\usepackage{listings}
\definecolor{lightgray}{gray}{1}

\usepackage{theorem}

\usepackage{tikz}

\newcommand{\nc}{\newcommand}
\nc{\braoprket}[3]{\langle#1|#2|#3\rangle}
\nc{\opn}[1]{\operatorname{#1}}
\nc{\avg}[1]{\langle#1\rangle}
\nc{\ketbrasame}[1]{|#1\rangle\!\langle#1|}
\nc{\swap}{\opn{SWAP}}
\nc{\E}{\mathbb{E}}
\nc{\Var}{\opn{Var}}
\nc{\dg}{\dagger}

\usepackage[normalem]{ulem}

\begin{document}

\title{Deconfined criticality as intrinsically gapless topological state in one dimension}

\author{Sheng Yang}
\affiliation{Institute for Advanced Study in Physics and School of Physics, Zhejiang University, Hangzhou 310058, China}

\author{Fu Xu}
\affiliation{Department of Physics, Nanjing University, Nanjing, Jiangsu 210093, China}

\author{Da-Chuan Lu}
\affiliation{Department of Physics and Center for Theory of Quantum Matter, University of Colorado,
Boulder, Colorado 80309, USA}
\affiliation{Department of Physics, Harvard University, Cambridge, MA 02138, USA}

\author{Yi-Zhuang You}
\email{yzyou@physics.ucsd.edu}
\affiliation{Department of Physics, University of California, San Diego, CA 92093, USA}

\author{Hai-Qing Lin}
\email{hqlin@zju.edu.cn}
\affiliation{Institute for Advanced Study in Physics and School of Physics, Zhejiang University, Hangzhou 310058, China}

\author{Xue-Jia Yu}
\email{xuejiayu@fzu.edu.cn}
\affiliation{Department of Physics, Fuzhou University, Fuzhou 350116, Fujian, China}
\affiliation{Fujian Key Laboratory of Quantum Information and Quantum Optics,
College of Physics and Information Engineering,
Fuzhou University, Fuzhou, Fujian 350108, China}

\begin{abstract}
Deconfined criticality and gapless topological states have recently attracted growing attention, as both phenomena go beyond the traditional Landau paradigm. However, the deep connection between these two critical states, particularly in lattice realization, remains insufficiently explored. In this Letter, we reveal that certain deconfined criticality can be regarded as an intrinsically gapless topological state without gapped counterparts in a one dimensional lattice model. Using a combination of field-theoretic arguments and large-scale numerical simulations, we establish the global phase diagram of the model, which features deconfined critical lines separating two distinct spontaneous symmetry breaking ordered phases. More importantly, we unambiguously demonstrate that the mixed anomaly inherent to deconfined criticality enforces topologically robust edge modes near the boundary, providing a general mechanism by which deconfined criticality manifests as a gapless topological state. Our findings not only offer a new perspective on deconfined criticality but also deepen our understanding of gapless topological phases of matter.

\end{abstract}

\maketitle

\emph{Introduction.}---Over the past two decades, the deconfined quantum critical point (DQCP) has attracted broad interest across the communities of condensed matter to high-energy physics, as it provides a paradigm for phase transitions that go beyond the Landau-Ginzburg-Wilson framework of symmetry breaking~\cite{Sachdev_1999,sondhi1997rmp,sachdev2023quantum,senthil2004deconfined,senthil2004prb,senthil2024deconfined,xu2012unconventional,Levin2004prb,senthil2004quantummattersphysicslandaus,senthil2006prb,Tanaka2005prl,Grover2008prl,Metlitski2008prb,Swingle2012prb,Moon2012prb,Slagle2014prb}. The DQCP deviates from the conventional Landau paradigm as its low-energy effective theory incorporates topological terms without classical counterparts, as well as associated quantum anomalies~\cite{senthil2006prb,Tanaka2005prl,SEIBERG2016395,Metlitski2016prb,Wang2017prx,Metlitski2017prb,You2018prx,Snir2018pnas,Jian2018prb,Bi2019prx,SENTHIL20191,Bi2020prr,Lee2019prx,song2019unifying,Song2020prx,zhang2025diracspinliquidunnecessary,Zhang2023prl}. Consequently, topology plays a crucial role in understanding DQCP, attracting significant efforts ranging from numerical simulations~\cite{Sandvik2007prl,Lou2009prb,Sandvik2010prl,Kaul2012prb,Kaul2008prb,Nahum2011prl,Melko2008prl,Block2013prl,Nahum2013prb,shao2016quantum,Suwa2016prb,DEmidio2017prl,Nahum2015prx,Nahum2015prl,Qin2017prx,Ma2018prb,Ma2019prl,li2019deconfinedquantumcriticalityemergent,Sreejith2019prl,Serna2019prb,Zhao2019NatPhys,Liu2019NC,Jiang2019prb,Roberts2019prb,Huang2019prb,Takahashi2020prr,Sandvik_2020,Roberts2021prb,Ogino2021prb_a,Ogino2021prb_b,Yang2023prb,Prembabu2024prb} to experimental realizations~\cite{Guo2020prl,Song2024NatPhys,Lee2023prl,cui2023proximate,guo2023deconfinedquantumcriticalpoint}. Despite these efforts, the nature of DQCP remains a topic of ongoing debate~\cite{Zhao2020prl,DEmidio2024prl,Wang2022SciPost,Ma2020prb,Nahum2020prb,Lu2021prb,Zhao2022prl,song2025evolutionentanglemententropysun,Chen2024prl,Zhou2024prx,chen2024emergentconformalsymmetrymulticritical,zhao2024scalingdisorderoperatorentanglement,Deng2024prl,Liu2024prl,Liu2022prx,LIU2024190,Christos2024prr,takahashi2024so5multicriticalitytwodimensionalquantum}, as recently highlighted by the discovery that the transition point in the SU(2) $J$-$Q$ spin model may correspond to a multicritical point with two relevant perturbations~\cite{Zhao2020prl,Lu2021prb,takahashi2024so5multicriticalitytwodimensionalquantum}, suggesting that revisiting DQCP is necessary and worthwhile.

\begin{figure}[!h]
    \includegraphics[width=0.8\linewidth]{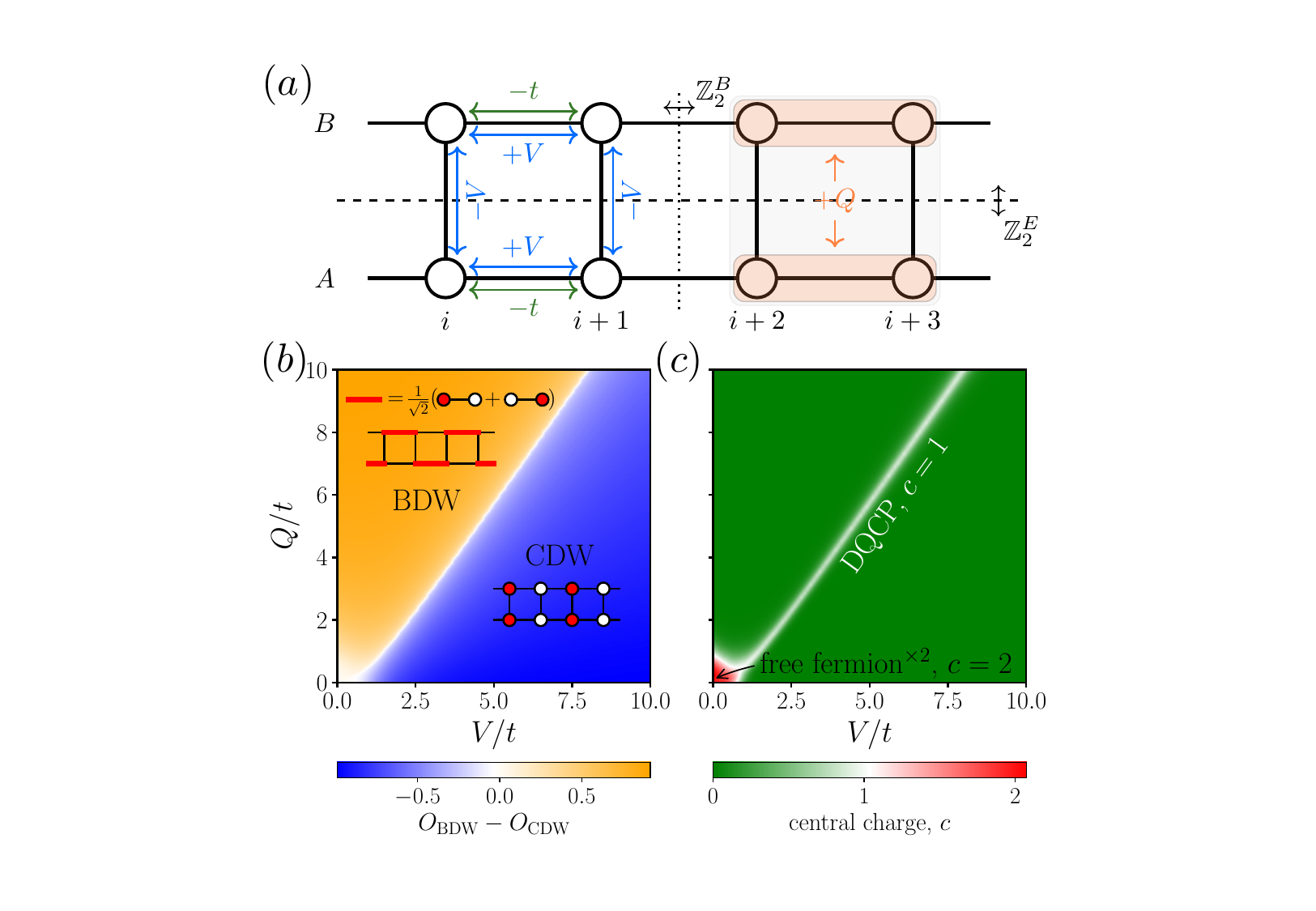}
    \caption{(a) Schematic plot of the two-leg spinless fermion ladder, where the green, blue, and orange lines represent the hopping ($t$), density-density ($V$), and bond-bond ($Q$) interaction terms in the Hamiltonian~\eqref{eq:hamiltonian}. $\mathbb{Z}_{2}^{B}$, and $\mathbb{Z}_{2}^{E}$ denote the  bond-centered reflection and layer-exchange symmetries, respectively. The ground-state phase diagram is mapped out numerically by (b) the local order parameters and (c) the estimated central charge, exhibiting BDW and CDW phases separated by a DQCP line. $O_\text{CDW/BDW}$ is obtained from infinite-size DMRG calculations with MPS bond dimension $\chi=200$, while the central charge $c$ is estimated by $6\times(S_{2}-S_{1})/(\ln\xi_{2}-\ln\xi_{1})$, where labels ``$1$'' and ``$2$'' indicate the bipartite entanglement entropy ($S$) and MPS correlation length ($\xi$) from $\chi=100$ and $200$ infinite-size DMRG simulations, respectively.}
    \label{fig:phasediagram}
\end{figure}

On a different front, recent advances~\cite{cheng2011prb,fidkowski2011prb,kestner2011prb,keselman2015prb,ruhman2017prb,parker2018prb,JIANG2018753,keselman2018prb,scaffidi2017prx,thorngren2021prb,verresen2021prx,verresen2020topologyedgestatessurvive,DuquePRB2021,yu2022prl,yu2024prl,parker2019prl,yu2024prb,yang2024giftslongrangeinteractionemergent,zhong2024pra,umberto2021sci_post,friedman2022prb,li2023intrinsicallypurelygaplesssptnoninvertibleduality,huang2023topologicalholographyquantumcriticality,wen2023prb,wen2023classification11dgaplesssymmetry,wen2024stringcondensationtopologicalholography,li2024sci_post,huang2024fermionicquantumcriticalitylens,su2024prb,zhang2024pra,ando2024gaugetheorymixedstate,zhou2024floquetenrichednontrivialtopologyquantum,li2024noninvertiblesymmetryenrichedquantumcritical,yu2025gaplesssymmetryprotectedtopologicalstates,tan2025exploringnontrivialtopologyquantum,zhong2025quantumentanglementfermionicgapless} show that many key features of topological physics persist even in the absence of a bulk energy gap, resulting in the concept of 
gapless symmetry-protected topological (gSPT) phases~\cite{scaffidi2017prx,verresen2021prx,thorngren2021prb}. A particularly intriguing class of gSPT phases, known as intrinsically gSPT phases~\cite{thorngren2021prb}, exhibits unique topological features arising from emergent anomalies that have no analogy in gapped systems. The discovery of gSPT states not only broadens the scope of topological physics to more challenging gapless systems but also provides new perspectives on the classification of phase transitions within the same universality class. This fundamentally enriches the conventional understanding of phase transitions, extending beyond the traditional Landau paradigm. 


In this Letter, we study a one-dimensional lattice model in which a DQCP coincides with an intrinsically gSPT state, as manifested by topological boundary modes protected by a mixed anomaly and coexisting with bulk critical fluctuations.
Using comprehensive investigation combining field theory arguments and large-scale numerical simulations, we establish the ground-state phase diagram of a newly designed interacting spinless fermion model on a two-leg ladder, which exhibits long-range charge-density wave (CDW) and bond-density wave (BDW) orders breaking incompatible $\mathbb{Z}_{2}$ symmetries. 
Furthermore, the BDW phase features topological edge modes protected by symmetry near the boundary and is therefore identified as a spontaneous SPT phase~\cite{Ma2022SciPost}. 
More importantly, a continuous phase transition line separating the spontaneous SPT and CDW phases exhibits intriguing critical and topological phenomena, as confirmed by numerical simulations and field-theoretic arguments. 
These findings are highlighted in two key aspects:
i) The transition line between the two $\mathbb{Z}_2$ symmetry breaking phases belongs to DQCP.
ii) The mixed anomaly associated with the DQCP protects the topological edge modes at criticality, which exist exclusively in gapless systems without gapped counterparts. 
Thus, we unambiguously demonstrate that the DQCP under study is an intrinsically gSPT phase, characterized by nontrivial boundary topological properties.

\emph{Model and method.}---We start by constructing the following microscopic model of interacting spinless fermions on a two-leg ladder of length $L$ at half filling [see Fig.~\ref{fig:phasediagram}(a)]
\begin{eqnarray}
\label{eq:hamiltonian}
    H &=& -\, t \sum_{i=1}^{L} \sum_{\alpha=A,B} D_{i,\alpha} + Q \sum_{i=1}^{L}  (D_{i,A}-1) (D_{i,B}-1) \nonumber  \\
    && +\, V \sum_{i=1}^{L} \sum_{\alpha=A,B} Z_{i,\alpha}Z_{i+1,\alpha} - V \sum_{i=1}^{L} Z_{i,A}Z_{i,B} \,.
\end{eqnarray}
The Hamiltonian is expressed purely in terms of bosonic operators $D_{i,\alpha} \equiv c^{\dagger}_{i,\alpha}c_{i+1,\alpha} + \text{h.c.}$ and $Z_{i,\alpha} \equiv c^{\dagger}_{i,\alpha}c_{i,\alpha}-1/2$, which are made of fermionic operators $c^{\dagger}_{i,\alpha}$ ($c_{i,\alpha}$) that create (annihilate) spinless fermions at rung $i$ on leg $\alpha$. These fermions will remain gapped throughout the phase diagram and will \emph{not} appear in the low-energy theory. They can be viewed as partons, only to construct a model for DQCP to happen \cite{2018PhRvB..98w5108I, 2024PhRvX..14b1044Z, 2024arXiv240504470C}. The UV lattice model has the anomaly-free symmetry $G_{UV}$ where the fermion parity $\mathbb{Z}_2^F$ (under which $c_{i,\alpha}\to-c_{i,\alpha}$) acts as a normal subgroup. In the low-energy theory, the fermion parity $\mathbb{Z}_2^F$ only acts on the gapped degrees of freedom, and the quotient symmetry $G_{UV}/\mathbb{Z}_2^F$ acts on the low-energy bosonic degrees of freedom.
In the model Hamiltonian $H$, $t$ is the fermion hopping amplitude and is set as the energy unit throughout, and 
$Q$ ($V$) represents the strength of the bond-bond (density-density) interactions. 
Besides the fermion pairity, $H$ respects the bond-centered reflection $\mathbb{Z}_{2}^{B}$, the layer-exchange $\mathbb{Z}_{2}^{E}$, and the layer fermion parity $\mathbb{Z}_{2}^{P}$ symmetries. In the low-energy, the quotient symmetry $G_{UV}/\mathbb{Z}_2^F\cong \mathbb{Z}_{2}^{B}\times \mathbb{Z}_{2}^{E} \times \mathbb{Z}_{2}^{P}$ have emergent mixed anomaly \cite{sal2024tdual,omer2025dqcp}, as the CDW domain wall will carry fractional charge of the unbroken symmetry.
These symmetries can be spontaneously broken by tuning the interaction parameters $Q/t$ and $V/t$, resulting in distinct long-range orders. 
In this work, we perform infinite- and finite-size density matrix renormalization group (DMRG) simulations~\cite{white1992prl,white1993prb,ulrich2011ap,mcculloch2008arxiv,stoudenmire2012arcmp} based on matrix product states (MPS)~\cite{verstraete2006prb,orus2014ap} to investigate bulk critical properties and boundary topological features of the model. 
A direct mapping of the model~\eqref{eq:hamiltonian} to a hard-core bosonic Hamiltonian via the standard Jordan-Wigner transformation is provided in Sec.~I of the Supplemental Material (SM) based on which the numerical simulations are performed.

\begin{figure*}[tb]
    \includegraphics[width=0.75\linewidth]{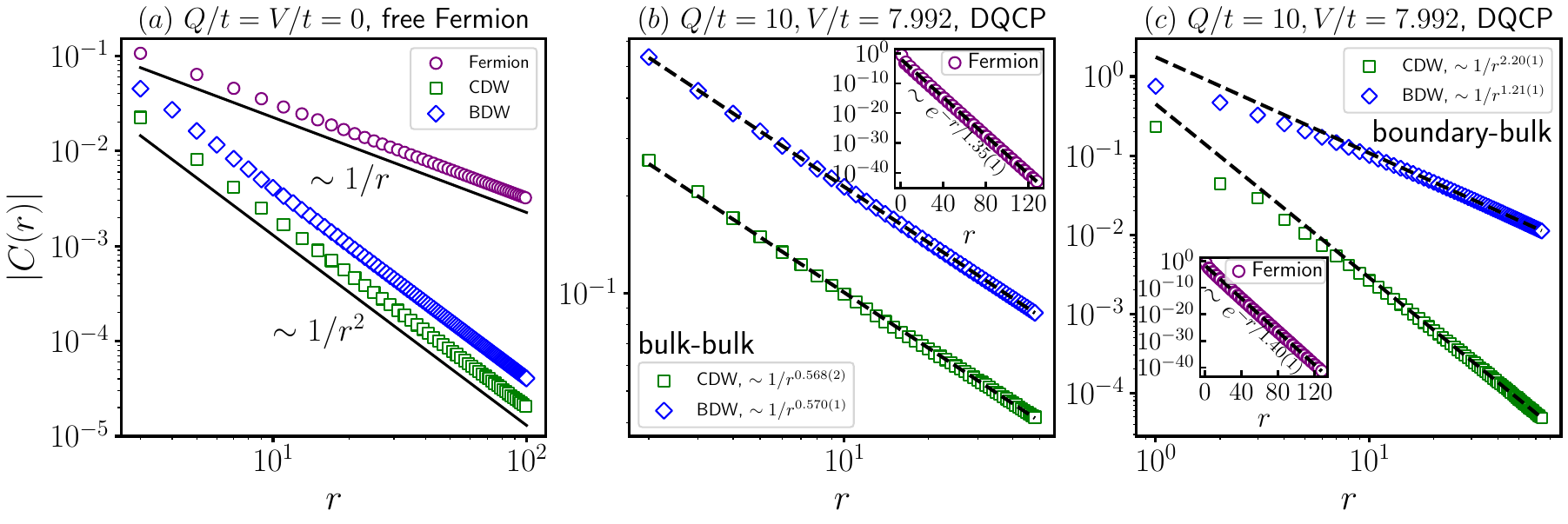}
    \caption{The connected correlation functions for fermion (purple circles), CDW (green squares), and BDW (blue diamonds) channels as a function of the distance $r$. (a) The scaling behaviors of the correlations at the exact free fermion point, $Q/t=V/t=0$, which can be computed analytically (see SM Sec.~IIIA). (b) The bulk-bulk correlations from infinite-size DMRG calculations at the estimated deconfined critical point, $V/t=7.992$, for $Q/t=10$. The data of CDW and BDW channels have been extrapolated to MPS bond $\chi\to\infty$ (see SM Sec.~IIIB). (c) The boundary-bulk correlations by putting one operator at the $1$th unit cell and the other one at the $(r+1)$th unit cell from finite-size DMRG simulations at the same critical point as (b). The data of CDW and BDW channels have been extrapolated to $L\to\infty$ to minimize the possible finite-size effect (see SM Sec.~IIIB). The dashed lines are least-squares fittings according to $a/r^{b}$ and $a e^{-r/\xi}$ in (b-c) and their insets, respectively.}
    \label{fig:correlations}
\end{figure*}

\emph{Quantum phase diagram.}---Before presenting the detailed DMRG results, we summarize our main findings and the ground-state phase diagram of the model. 
For $V/t \gg Q/t$, the ground state is an insulating phase with long-range CDW order, which spontaneously breaks the bond-centered reflection symmetry $\mathbb{Z}_{2}^{B}$. 
Conversely, for $Q/t \gg V/t$, the bond-bond interaction dominates, leading to the long-range BDW order, which breaks the layer-exchange $\mathbb{Z}_{2}^{E}$ symmetries. 
In the presence of competing quantum fluctuations from density-density and bond-bond interactions, the global phase diagram is numerically mapped out as shown in Fig.~\ref{fig:phasediagram}(b). 
The diagram reveals distinct CDW and BDW ordered phases separated by a continuous phase transition line. 
Remarkably, the BDW phase not only exhibits long-range order but also features dangling fermion modes near the open boundaries. 
More importantly, we unambiguously demonstrate the existence of four fixed points or lines in the phase diagram [see Fig.~\ref{fig:phasediagram}(c)]. 
Two of them are unstable with central charges $c= 1$ and $2$, corresponding to a continuous deconfined phase transition between the two ordered phases and a two-copy free fermion point at $Q/t = V/t = 0$, respectively. 
The other two are stable with $c=0$, corresponding to stable phases with distinct long-range orders. 
Intriguingly, in the following sections, we will present a comprehensive argument that the entire critical line with $c=1$ represents a special type of DQCP with topologically protected edge modes, which share key features with intrinsically gSPT phases.

\emph{Spontaneous SPT phase and deconfined criticality.}---Unlike conventional SPT phases that preserve all symmetries, the BDW phases exhibit a coexistence of symmetry-breaking and topological edge modes, referred to as spontaneous SPT phases~\cite{Ma2022SciPost}. 
The degeneracy of the BDW phase under the open boundary condition is $2 \times 2 = 4$, where the two factors $2$ come from spontaneous symmetry breaking and edge modes near the boundary, respectively. 
As detailed in Sec.~II of the SM, in BDW phases, spontaneous symmetry breaking is characterized by the long-range order of the BDW order parameter $O_\text{BDW} \sim (-1)^{i} (D_{i,A} - D_{i,B})$, while the nontrivial topological edge modes can be directly reflected in the local density distribution and evidenced by a nontrivial degeneracy in the bulk entanglement spectrum~\cite{Li2008prl,Pollmann2010prb}. 
In fact, this special type of SPT phase is protected by the bond-centered reflection $\mathbb {Z}_{2}^{B}$ symmetry.



We now turn to the transition line in the phase diagram that separates two incompatible symmetry-breaking ordered phases. 
As a first step, we analytically calculate various types of connected bulk-bulk correlation functions for $Q/t = V/t = 0$ (see SM Sec.~IIIA). 
As shown in Fig.\ref{fig:correlations}(a), it reveals that all connected correlation functions exhibit algebraic decay, with the fermion scaling dimension $\Delta_{F} = 1/2$. 
This value is half that of both order parameters, $\Delta_\text{BDW/CDW} = 1$, consistent with the $c=2$ two-copy free fermion fixed point. 
Furthermore, upon including the density-density and bond-bond interaction terms, it is natural to ask whether a DQCP may emerge by tuning $V/t$ and $Q/t$. 
To address this issue, we demonstrate the continuous phase transition (excluding fine-tuned cases) between the two ordered phases, as evidenced by both order parameters vanishing continuously at a single point (see SM Sec.~II). 
We further accurately determined the critical point and critical exponents through finite-length scaling analyses (see SM Sec.~II). 
At these critical points, we find that both order parameters unambiguously demonstrate the expected power-law behavior and identical scaling dimensions, $\Delta_\text{BDW} \approx \Delta_\text{CDW}$ [also see Fig.~\ref{fig:correlations}(b) for the results of connected correlation functions]. 
These results support an emergent $\mathrm{U}(1)_\theta$ symmetry at the criticality, a hallmark feature of DQCP. In fact, the critical point is described by a compact boson conformal field theory (CFT) at central charge $c=1$ with $\mathrm{U}(1)_\theta\times\mathrm{U}(1)_\phi$ symmetry~\cite{Huang2019prb}\footnote{At the DQCP, both the $\mathrm{U}(1)_\theta$ and $\mathrm{U}(1)_\phi$ symmetries emerge, corresponding to the momentum and winding symmetries in the $c=1$ compact boson CFT related by T duality, where the BDW-CDW rotation symmetry is only one of these $\mathrm{U}(1)$ symmetries.}, which serves as the low-energy effective theory of the (1+1)D DQCP.

\begin{figure}[tb]
    \includegraphics[width=1.0\linewidth]{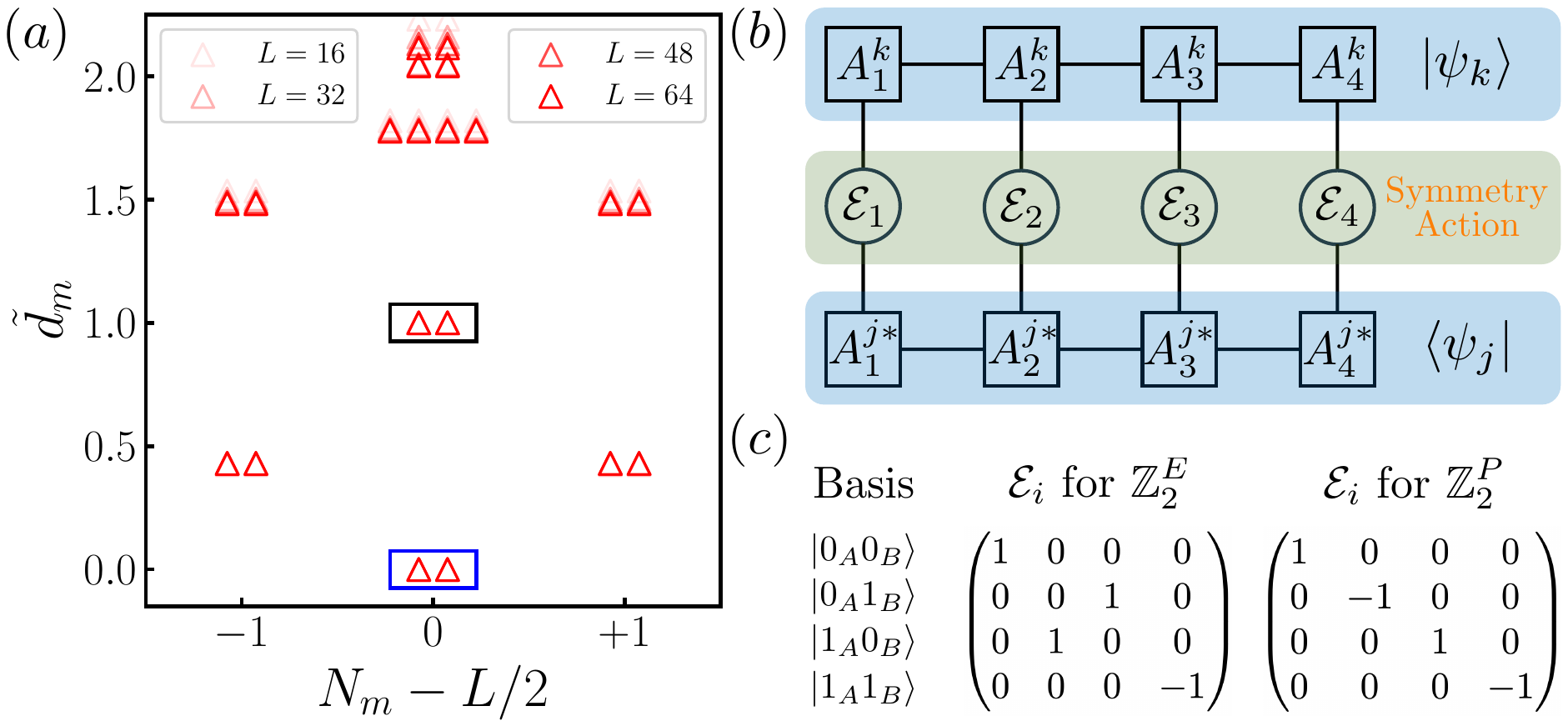}
    \caption{(a) The entanglement spectrum $d_{m} = -\log{\lambda_{m}}$ where $\lambda_{m}$ is the eigenvalue of the half-chain reduced density matrix $\rho_{\rm r}$ calculated for $L=16$ to $64$ under periodic boundary conditions at the estimated critical point $V/t=7.992$ for $Q/t=10$. $\tilde{d}_{m}$ is the rescaled spectrum after fixing the two framed levels to $0$ and $1$, respectively. $N_{m}$ is the occupation number of the left half chain $\sum_{i=1}^{L/2}\sum_{\alpha} \bra{\phi_{m}}  n_{i,\alpha} \ket{\phi_{m}}$ associated to $\lambda_m$ and $\ket{\phi_{m}}$ is the corresponding eigenstate of $\rho_\text{r}$. (b) The tensor network diagram for the representation of the symmetry actions, the layer exchange $\mathbb{Z}_{2}^{E}$ and the layer fermion parity $\mathbb{Z}_{2}^{P}$, within a two-fold degenerate subspace spanned by $\{ |\psi_{0}\rangle, |\psi_{1}\rangle \}$. For global on-site symmetries, the generator has the form $\prod_{i=1}^{L} \mathcal{E}_{i}$. (c) gives the concrete matrix form of the on-site operator $\mathcal{E}_{i}$ for $\mathbb{Z}_{2}^{E}$ and $\mathbb{Z}_{2}^{P}$ in the local computational basis introduced in SM Sec.~I.}
    \label{fig:es_rep}
\end{figure}

\emph{Topological edge modes at the DQCP.}---Recent advances~\cite{verresen2018prl,verresen2021prx,yu2022prl} suggest the existence of topological classifications of quantum critical points, characterized by the coexistence of nontrivial topological edge modes and bulk fluctuations under open boundary conditions. 
To verify this at the DQCP, we examine the connected correlation functions at the critical point $Q/t=10$ and $V/t=7.992$, as shown in Figs.~\ref{fig:correlations}(b) and~(c). 
By investigating the bulk-bulk correlations, we can see that the fermion degrees of freedom are gapped in the bulk and there is an U(1)$_{\theta}$ symmetry emerged at the critical point rotating between BDW and CDW orders. The nontrivial topology at criticality is reflected in the bulk entanglement spectrum~\cite{yu2024prl,zhang2024pra,zhong2025quantumentanglementfermionicgapless}. As shown in Fig.~\ref{fig:es_rep}(a), the numerical results show that as the system size increases, the lowest level of the bulk entanglement spectrum at criticality encodes a robust two-fold topological degeneracy (see SM Sec.~IV for additional numerical evidence), implying the existence of topologically nontrivial fermion modes near the boundaries. These fermionic modes lead to nontrivial boundary criticality, manifesting in distinct boundary scaling dimensions for both order parameters, as evidenced by boundary-bulk correlations in Fig.~\ref{fig:correlations}(b). Finally, the exponential decay of the fermion boundary-bulk correlation suggests the existence of gapless fermionic edge modes that are exponentially localized near the boundary.

\emph{DQCP as an intrinsically gSPT.}---To uncover the deep connection between DQCP and the intrinsically gSPT phase, we address two fundamental questions: i) How can we demonstrate the existence of gapless fermion boundary modes at the DQCP? ii) How can we show that these fermion boundary modes are unique to gapless systems without gapped counterparts?

We address the first question by examining the mixed anomalies between the $\mathbb{Z}_{2}$ symmetries associated with the distinct ordered phases. 
Specifically, we deform the Hamiltonian~\eqref{eq:hamiltonian} with additional pinning fields on the boundary, $-h \sum_{\alpha=A,B} (n_{1,\alpha} + n_{L,\alpha})$ where $h/t=10$ the same order as $Q/t$, to construct a CDW domain wall at criticality. 
The ground state of the deformed Hamiltonian exhibits a twofold degeneracy (a visualization of the two states can be found in SM Sec.~IV). 
In the subspace of the two-fold degenerate states at the critical point, the layer fermion parity $\mathbb{Z}_{2}^{P}$ and the layer-exchange $\mathbb{Z}_{2}^{E}$ symmetry operations can be represented as the effective $\sigma_{z}$ and $\sigma_{x}$ operators, respectively, which anticommute with each other. 
Fig.~\ref{fig:es_rep}(b) and (c) provide the tensor-network diagram for the symmetry analysis within the two-fold subspace. 
This indicates that the symmetry acts as the projective representation of the $\mathbb{Z}_{2}^{E} \times \mathbb{Z}_{2}^{P}$. 
In other words, the CDW domain wall carries the fractional charge of the BDW order, consistent with the mixed anomalies between the $\mathbb{Z}_{2}^B$ and $\mathbb{Z}_{2}^P \times \mathbb{Z}_2^E$ symmetries at the DQCP~\cite{Wang2017prx,Huang2019prb,Ma2022SciPost}. 
This projective representation can only be realized by gapless fermions trapped at the CDW domain wall, which must originate from the boundary because the bulk fermions are gapped [see the inset of Fig.~\ref{fig:correlations}(b)]. 
Therefore, the mixed anomalies enforce the presence of gapless fermions near the boundary at the DQCP.

To address the second question, it is well known that realizing gapless boundary fermions in a one-dimensional gapped system requires the system to be in a fermionic SPT phase, such as the Su-Schrieffer-Heeger model. 
In this context, realizing nontrivial topological edge states requires breaking at least one $\mathbb{Z}_{2}$ symmetry, regardless of the dimerization arrangement. 
However, at the DQCP, all global symmetries are preserved. 
Thus, it is fundamentally impossible to realize the above topological edge states in any gapped system. 
This demonstrates that the gapless boundary fermions observed at the DQCP are unique to gapless systems and cannot be replicated in gapped counterparts.

\emph{Concluding remarks.}---To summarize, we combine large-scale numerical simulations and field theory arguments to uncover the nontrivial boundary physics of the one-dimensional DQCP and demonstrate that it can be regarded as an intrinsically gSPT phase. 
Utilizing a newly designed spinless fermion model with competing density-density and bond-bond interactions on a two-leg ladder, we decipher the ground-state phase diagram, which exhibits continuous phase transition separating CDW and BDW long-range ordered phases, each of which breaks distinct $\mathbb{Z}_{2}$ symmetries. 
The BDW phases are identified as spontaneous SPT phases, characterized by both spontaneous symmetry breaking and symmetry-protected topological edge modes. 
More importantly, supported by numerical simulations and field-theoretic arguments, we unambiguously demonstrate that the transition points between the spontaneous SPT and CDW long-range ordered phases not only correspond to DQCP but also feature topological edge modes without gapped counterparts, which are attributed to the mixed anomaly inherent to the DQCP.

Regarding experimental realization, we note that the spinless fermion ladder model can be implemented using ultracold atom quantum simulators. 
Additionally, in two dimensions, the deconfined quantum phase transition between a quantum spin Hall insulator and an $s$-wave superconducting phase also features an intrinsically gSPT phase~\cite{Ma2022SciPost}, which could potentially be realized experimentally in materials such as WTe$_2$~\cite{Liu2019NC,Song2024NatPhys}. 
Our work not only reveals a new nature of the deconfined quantum criticality from the perspective of boundary physics, but also contributes to a deeper understanding of gapless topological phases of matter.

\textit{Acknowledgement}: 
Numerical simulations were carried out with the ITENSOR \verb|C++| package~\cite{itensor}. X.-J. Yu was supported by the National Natural Science Foundation of China (Grant No.12405034) and a start-up grant from Fuzhou University. Y.Z.Y. is supported by the NSF Grant No. DMR-2238360. Research of D.C.L. after September 2024 is supported by the Simons Collaboration on Ultra-Quantum Matter, which is a grant from the Simons Foundation 651440. The work of S.Y. is supported by China Postdoctoral Science Foundation (Certificate Number: 2024M752760). This work is also supported by MOST 2022YFA1402701.

\bibliographystyle{apsrev4-2}
\let\oldaddcontentsline\addcontentsline
\renewcommand{\addcontentsline}[3]{}
\bibliography{main.bib}

\let\addcontentsline\oldaddcontentsline
\onecolumngrid

\clearpage
\newpage

\widetext

\begin{center}
\textbf{\large Supplemental Material for ``Deconfined criticality as intrinsically gapless topological state in one dimension''}
\end{center}

\maketitle

\renewcommand{\thefigure}{S\arabic{figure}}
\setcounter{figure}{0}
\renewcommand{\theequation}{S\arabic{equation}}
\setcounter{equation}{0}
\renewcommand{\thesection}{\Roman{section}}
\setcounter{section}{0}
\setcounter{secnumdepth}{4}

\addtocontents{toc}{\protect\setcounter{tocdepth}{0}}
{
\tableofcontents
}

\section{Map the two-leg spinless fermion ladder model to a hard-core boson model}
\label{sm:mapping}

In order to study the bulk and boundary physics of the two-leg ladder spinless fermion model introduced in the main text, we have performed extensive finite- and infinite-size density matrix renormalization group (DMRG)~\cite{white1992prl,white1993prb} simulations based on matrix product state (MPS)~\cite{verstraete2006prb,orus2014ap} representations.

To facilitate the numerical calculations, we first group each two sites on the same rung (e.g., the $i$th rung) as a single composite site, whose Hilbert space is spanned by $\{ \ket{0_{A}0_{B}}, \ket{0_{A}1_{B}}, \ket{1_{A}0_{B}}, \ket{1_{A}1_{B}} \}$ of local dimension $d=4$ where $0$ and $1$ denote the occupation number of fermions. 
Then we use the standard Jordan-Wigner transformation to map the spinless fermions ($c$ and $c^{\dagger}$ operators) to hard-core bosons ($a$ and $a^{\dagger}$ operators),
\begin{eqnarray}
    c_{i,A} &=& F_{1} F_{2} \cdots F_{i-1} a_{i,A} \,, \\
    c_{i,B} &=& F_{1} F_{2} \cdots F_{i-1} (F_{i} a_{i,B}) \,,
\end{eqnarray}
where $F_{j} \equiv (1-2n_{j,A})(1-2n_{j,B}) = (-1)^{n_{j,A}+n_{j,B}}$ is the fermion phase operator at the $j$th site.
Using this map, the model Hamiltonian can be easily rewritten by bosonic operators,
\begin{align}
  H_{\rm F \to B} = & - (t+Q) \sum_{i} \Big[ (a^{\dagger}_{i,A}F_{i}) a_{i+1,A} + (F_{i}a_{i,A}) a^{\dagger}_{i+1,A} + a^{\dagger}_{i,B} (F_{i+1}a_{i+1,B}) + a_{i,B} (a^{\dagger}_{i+1,B}F_{i+1}) \Big] \nonumber \\
  & + Q \sum_{i} \Big[ (a^{\dagger}_{i,A}a^{\dagger}_{i,B}F_{i}) (F_{i+1}a_{i+1,A}a_{i+1,B}) - (a^{\dagger}_{i,A}F_{i}a_{i,B}) (a^{\dagger}_{i+1,B}F_{i+1}a_{i+1,A}) \Big] \nonumber \\
  & + Q \sum_{i} \Big[ - (a^{\dagger}_{i,B}F_{i}a_{i,A}) (a^{\dagger}_{i+1,A}F_{i+1}a_{i+1,B}) + (F_{i}a_{i,A}a_{i,B}) (a^{\dagger}_{i+1,A}a^{\dagger}_{i+1,B}F_{i+1}) \Big] \nonumber \\
  & + V \sum_{i} (Z_{i,A}Z_{i+1,A} + Z_{i,B}Z_{i+1,B} - F_{i}/4)  + \text{const} \,.
\end{align}
Here, $Z_{i,\alpha} \equiv n_{i,\alpha} - 1/2$ and operators defined on the same composite site have been grouped together by explicit parentheses. 
Having transformed the spinless fermion Hamiltonian into the above form, standard finite- and infinite-size DMRG calculations can be performed following conventional literature~\cite{ulrich2011ap,mcculloch2008arxiv,stoudenmire2012arcmp}.

\section{Deconfined quantum criticality and finite-length scaling analysis}
\label{sm:dqcp}

To determine the precise location of the deconfined quantum critical point and investigate its universal behaviors, we perform extensive infinite-size DMRG simulations and apply finite-length scaling analyses. 
Without loss of generality, we fix $Q/t = 10$ and $V/t$ is the remaining tuning parameter in the model. 
As both the relevant long-range orders, i.e., the charge-density wave (CDW) and bond-density wave (BDW) orders, break the one-site translation symmetry explicitly, the period of the repeating unit of the infinite MPS should be even, which is four in our work. 

\begin{figure*}[tb]
    \includegraphics[width=0.85\linewidth]{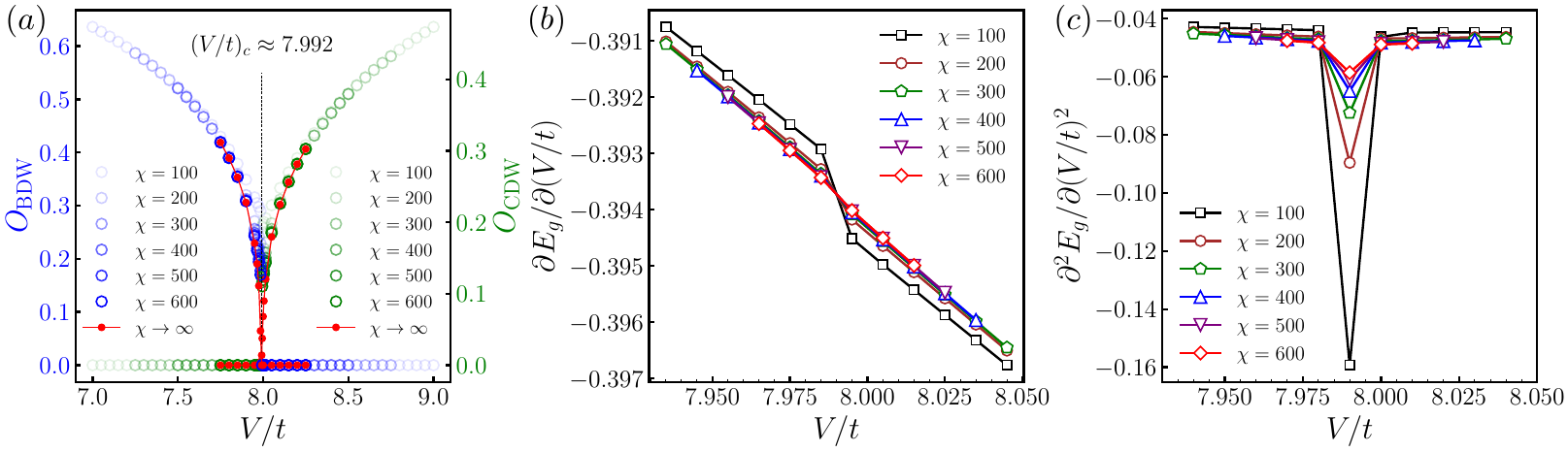}
    \caption{(a) Order parameters of the BDW and CDW phases as a function of the tuning parameter $V/t$ along the fixed $Q/t=10$ line for different MPS bond dimensions. The red solid circles are obtained by extrapolating finite-$\chi$ data to the limit $\chi \to \infty$ according to $O_\text{BDW/CDW}(V/t, \chi) = a (100/\chi)^{b} + O_\text{BDW/CDW}(V/t, \chi \to \infty)$ for each $V/t$. (b) The first-order and (c) the second-order derivatives of the ground-state energy per rung as a function of $V/t$ with $Q/t=10$ for various bond dimensions. Numerical data are obtained from infinite-size DMRG calculations.}
    \label{fig:idmrg_order_energy}
\end{figure*}

To detect the long-range CDW and BDW orders, we first calculate the corresponding local order parameters
\begin{eqnarray}
    O_\text{CDW} &\sim& (-1)^{i} (Z_{i,A}+Z_{i,B}) \,, \\
    O_\text{BDW} &\sim& (-1)^{i} (D_{i,A} - D_{i,B}) \,,
\end{eqnarray}
where $Z_{i,\alpha} \equiv n_{i,\alpha}-1/2$ and $D_{i,\alpha} \equiv c_{i,\alpha}^{\dagger}c_{i+1,\alpha} + c_{i+1,\alpha}^{\dagger}c_{i,\alpha}\,$. 
As shown in Fig.~\ref{fig:idmrg_order_energy}(a), by averaging within the repeating unit and then extrapolating to infinite bond dimension $\chi \to \infty$ for each $V/t$, both order parameters $O_\text{CDW}$ and $O_\text{BDW}$ continuously vanish from two sides at a single point $(V/t)_{c}$, indicating a direct continuous phase transition between two long-range orders. 
Furthermore, Figs.~\ref{fig:idmrg_order_energy}(b) and (c) demonstrate the first- and second-order derivatives of the ground-state energy $E_{g}$ versus the tuning parameter $V/t$ for different bond dimensions $\chi$. 
The first-order derivative $\partial E_{g} / \partial (V/t)$ is found to become continuous as $\chi$ increases, while the second-order derivative $\partial^{2} E_{g} / \partial (V/t)^{2}$ exhibits a singularity near the critical point $(V/t)_{c} \approx 7.992$\,. 
We also note that the singularity shown in $\partial^{2} E_{g} / \partial (V/t)^{2}$ decreases with larger $\chi$, which implies that the transition might be higher order.

\begin{figure*}[tb]
    \includegraphics[width=0.7\linewidth]{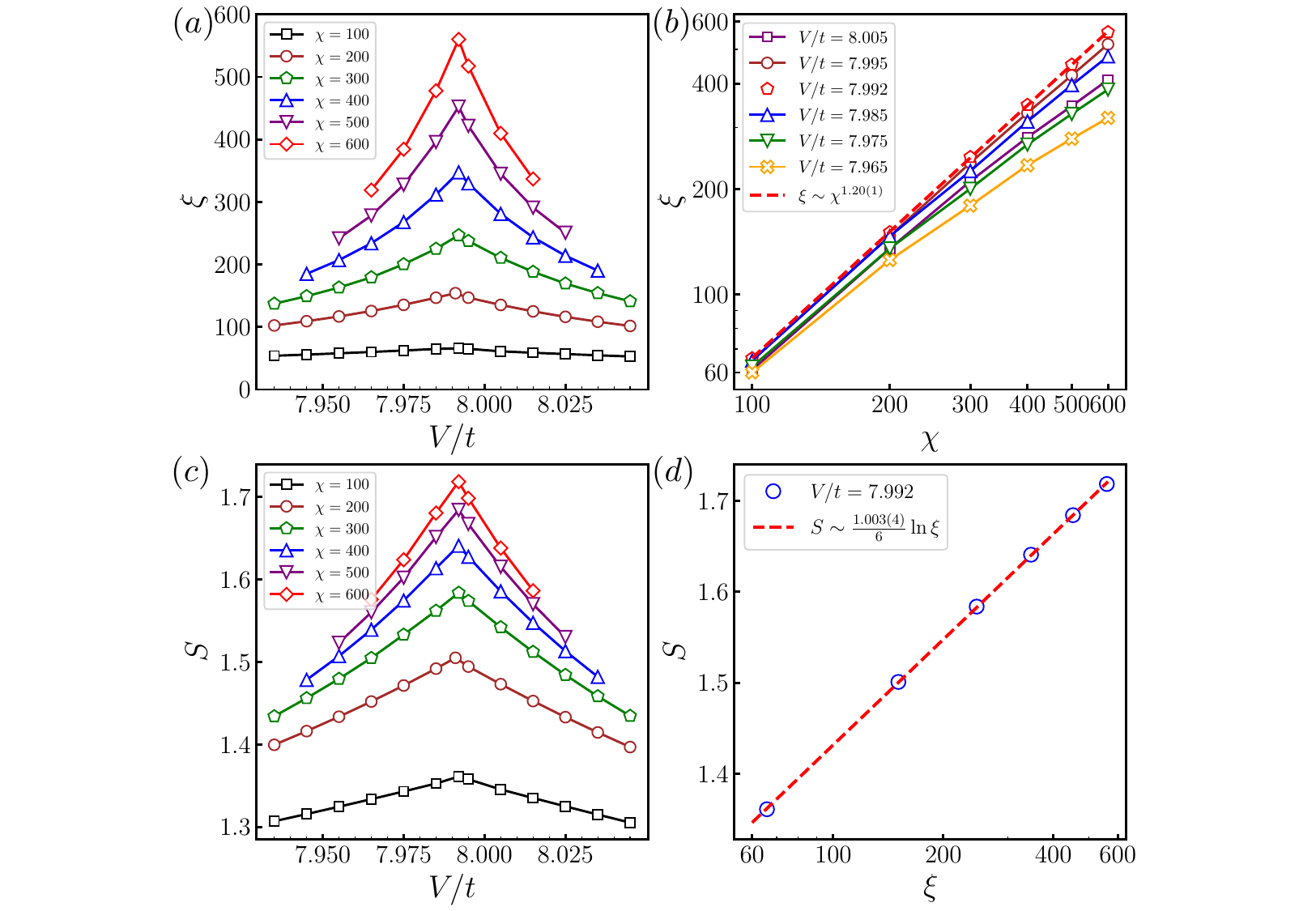}
    \caption{(a) The effective correlation length $\xi$ of the MPS and (c) the bipartite entanglement entropy $S$ as a function of the driving parameter $V/t$ with fixed $Q/t=10$ near the critical point for bond dimensions $\chi$ from $100$ to $600$. (b) A log-log plot of $\xi$ versus $\chi$ for various $V/t$ near the critical point. $\xi$ displays a power-law dependence on $\chi$ at the estimated critical point $(V/t)_{c} = 7.992$; the red dashed line, $\xi \sim \chi^{1.20(1)}$, is a least-squares fitting. (d) The entanglement entropy from different $\chi$ as a function of $\xi$ at the critical point $(V/t)_{c} = 7.992$\,. A least-squares fitting according to $S \sim c/6 \log\xi$ estimates the central charge $c=1.003(4)$\,. Numerical data are obtained from infinite-size DMRG calculations.}
    \label{fig:idmrg_xi_entropy}
\end{figure*}

In order to further reveal the characteristic features of the quantum criticality, we investigate the performances of the bipartite entanglement entropy $S$ and the MPS correlation length $\xi$ across the critical point estimated above. 
The bipartite entanglement entropy can be accessed directly within the infinite-size DMRG simulation process using the Schmidt coefficients $\lambda_{i}$ through $S = - \sum_{i=1}^{\chi} \lambda_{i}^{2} \log{\lambda_{i}^{2}}$, while the effective correlation length $\xi$ of the MPS is evaluated by $\xi = -1/\log(|\tau_{2}/\tau_{1}|)$ where $\tau_{1(2)}$ is the (second) largest eigenvalue of the so-called transfer matrix~\cite{ulrich2011ap,orus2014ap}. 
It is noted that the effective correlation length $\xi$ serves as an intrinsic property of the infinite MPS that determines the characteristic distance over which correlations can propagate. 
As shown in Figs.~\ref{fig:idmrg_xi_entropy}(a) and (c), it is clear that both $S$ and $\xi$ exhibit a divergence at $(V/t)_{c} \approx 7.992$ as expected. 
Furthermore, we can see that $\xi$ right at the critical point displays a power law behavior with $\chi$ described by $\xi \sim \chi^{\kappa}$ where $\kappa = 1.20(1)$ [refer to Fig.~\ref{fig:idmrg_xi_entropy}(b)], which is supported by the finite-entanglement scaling theory~\cite{fes2008prb,fes2009prl,fes2012prb,fes2014prb}. 
To further extract the central charge of the underlying conformal field theory, we also plot $S$ as a function of $\xi$ at the estimated critical point as displayed in Fig.~\ref{fig:idmrg_xi_entropy}(d). 
A least-squares fitting according to the formula, $S \sim \frac{c}{6} \log\xi$~\cite{calabrese2004jsm,calabrese2009jpa}, gives $c = 1.003(4)$\,. 

\begin{figure*}[tb]
    \includegraphics[width=0.85\linewidth]{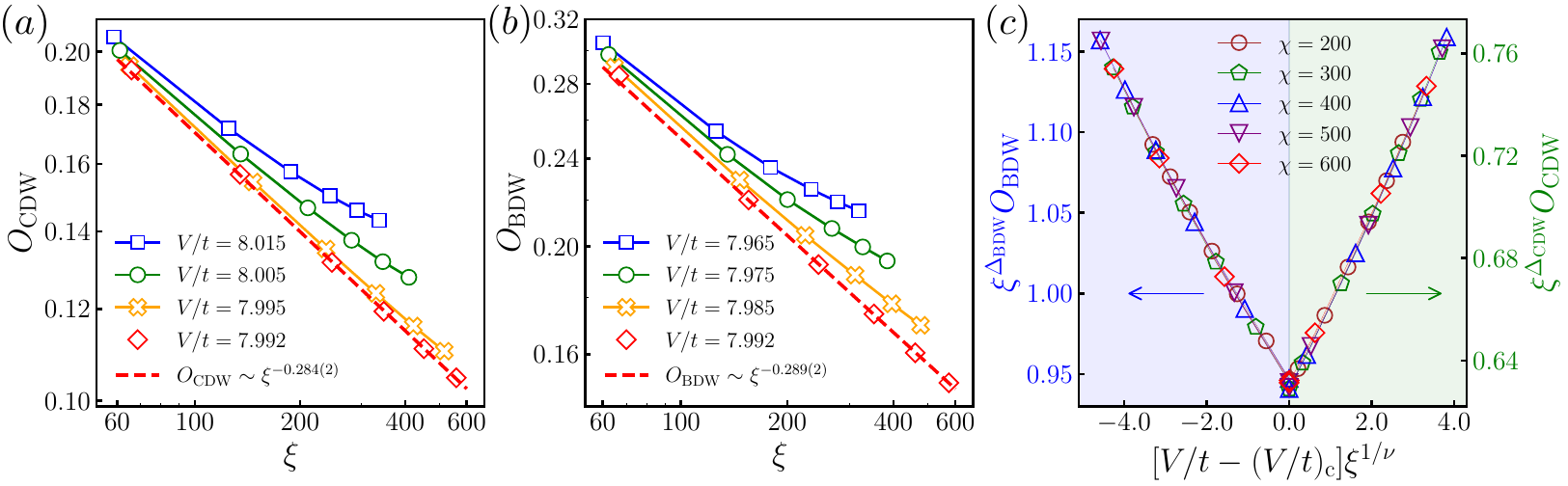}
    \caption{Order parameters of the (a) CDW and (b) BDW phases as a function of the effective correlation length $\xi$ near the critical point for $Q/t=10$. $O_{\rm BDW/CDW}$ shows perfect finite-length scaling behaviors obeying $O_{\rm BDW/CDW} \sim \xi^{-\Delta_{\rm BDW/CDW}}$ at $V/t=7.992$; Standard least-squares fittings give the estimations $\Delta_{\rm CDW} = 0.284(2)$ and $\Delta_{\rm BDW} = 0.289(2)$\,. (c) Data collapse of the order parameters of different MPS bond dimensions and tuning parameters close to the critical point. By using the scaling dimensions extracted from (a) and (b), rescaled data points $O_{\rm BDW/CDW}\xi^{\Delta_{\rm BDW/CDW}}$ can fall onto a single curve as a function of $(V/t-7.992)\xi^{1/\nu}$ with $\nu = 1.14(4)$ for BDW and $\nu = 1.17(5)$ for CDW. Numerical data are obtained from infinite-size DMRG calculations.}
    \label{fig:idmrg_scaling}
\end{figure*}

Finally, it is also important to obtain the critical exponents describing the universal behaviors of the phase transition. 
This task can be accomplished by performing finite-length scaling analyses. 
For infinite MPS simulations, the effective correlation length $\xi$ is usually seen as the finite length scale that enters the scaling relations. 
According to the scaling theory, order parameters near the critical point should obey the following universal form~\cite{nigel1992book}
\begin{equation}
    \label{eq:scaling}
    O_\text{BDW/CDW}(V/t,\xi) = \xi^{-\Delta_\text{BDW/CDW}} \mathcal{G}[(V/t - (V/t)_{c}) \xi^{1/\nu}] \,,
\end{equation}
where $\Delta_\text{BDW(CDW)}$ is the scaling dimension of the BDW (CDW) order parameter, $\nu$ is the correlation length exponent and $\mathcal{G}$ is an unknown universal function. 
In particular, right at the critical point, it is expected that order parameters should display power law behaviors, $O_\text{BDW/CDW} \sim \xi^{-\Delta_\text{BDW/CDW}}$, according to Eq.~\eqref{eq:scaling}. 
As shown in Fig.~\ref{fig:idmrg_scaling}(a) and (b), this power law behavior is observed for both order parameters exactly at the estimated critical point $(V/t)_{c} = 7.992$ with fitted exponents $\Delta_\text{BDW} = 0.289(2)$ and $\Delta_\text{CDW} = 0.284(2)$\,. 
The identical scaling dimensions imply an emergent $O(2)$ symmetry at the critical point, which is a hallmark feature of the deconfined criticality. 
After substituting the estimated values of the critical point and the scaling dimensions into Eq.~\eqref{eq:scaling}, $\nu$ becomes the only parameter we need to adjust in the finite-length scaling analysis. 
As shown in Fig.~\ref{fig:idmrg_scaling}(c), a perfect data collapse of the order parameters from different $\chi$ can be achieved near the critical point with $\nu = 1.14(4)$ for the BDW order and with $\nu = 1.17(5)$ for the CDW order, respectively. 
This result ends the basic analysis of the deconfined criticality at the transition point between BDW and CDW ordered phases.

\section{Connected correlation functions at the free fermion point and the deconfined quantum critical point}

In this section, we provide the derivation of the connected correlation functions for the fermion, CDW, and BDW channels at the exact free fermion point (i.e., $Q/t = V/t = 0$), and explain the finite-bond and finite-size extrapolations used in Figs.~2(b) and (c) in the main text.

\subsection{Analytic expressions of the connected correlation functions at the free fermion point}

For the exact free fermion point, i.e., $Q/t = V/t = 0$, the two-leg ladder Hamiltonian is reduced to two decoupled free fermion chains (periodic boundary condition is assumed here)
\begin{equation}
    H_\text{FF} = - t \sum_{j=1}^{L} \sum_{\alpha=A,B} \left( c_{j,\alpha}^{\dagger} c_{j+1,\alpha}^{} + c_{j+1,\alpha}^{\dagger} c_{j,\alpha}^{} \right) \,.
\end{equation}
After the standard Fourier transformation, $c_{k,\alpha} \equiv \frac{1}{\sqrt{L}} \sum_{j=1}^{L} e^{-\text{i}jk} c_{j,\alpha}$, the Hamiltonian becomes 
\begin{equation}
    H_\text{FF} = -2t \sum_{\alpha,k} \cos(k) c_{k,\alpha}^{\dagger} c_{k,\alpha}^{} \,,
\end{equation}
where $k = 2\pi m/L$ with $m = -L/2+1, \cdots, L/2$.

Now the fermion-fermion connected correlation function with respect to the ground state of $H_\text{FF}$ at half filling can be calculated by (here, $r \equiv j-i$)
\begin{eqnarray}
    2 C_\text{Fermion}(r) &\equiv& \sum_{\alpha} \left[ \langle c_{i,\alpha}^{\dagger} c_{j,\alpha}^{} \rangle - \langle c_{i,\alpha}^{\dagger} \rangle \langle c_{j,\alpha} \rangle \right] = \frac{1}{L} \sum_{\alpha} \sum_{k,k'} e^{-\text{i}ik + \text{i} jk'} \langle c_{k,\alpha}^{\dagger} c_{k',\alpha}^{} \rangle \nonumber \\
    &=& \frac{1}{L} \sum_{\alpha} \sum_{k} e^{\text{i} kr} \langle c_{k,\alpha}^{\dagger} c_{k,\alpha}^{} \rangle \xrightarrow{L \to \infty} \frac{1}{\pi} \int_{-\pi/2}^{\pi/2} e^{\text{i} kr} \text{d}k = \frac{2 \sin(\pi r/2)}{\pi r} \,.
\end{eqnarray}

Similarly, we can obtain the connected correlation functions for the CDW and BDW channels as follows:
\begin{eqnarray}
    C_\text{CDW}(r) &\equiv& (-1)^{i+j} \langle (Z_{i,A} + Z_{i,B}) (Z_{j,A} + Z_{j,B}) \rangle - (-1)^{i+j} \langle (Z_{i,A} + Z_{i,B}) \rangle \langle (Z_{j,A} + Z_{j,B}) \rangle \nonumber \\
    &=& (-1)^{i+j} \left[ \langle c_{i,A}^{\dagger} c_{j,A} \rangle \langle c_{i,A} c_{j,A}^{\dagger} \rangle + \langle c_{i,B}^{\dagger} c_{j,B} \rangle \langle c_{i,B} c_{j,B}^{\dagger} \rangle \right] \nonumber \\
    &=& (-1)^{r+1} \frac{2\sin^{2}(\pi r/2)}{(\pi r)^{2}} \,,
\end{eqnarray}
and
\begin{eqnarray}
    C_\text{BDW}(r) &\equiv& (-1)^{i+j} \langle (D_{i,A} - D_{i,B}) (D_{j,A} - D_{j,B}) \rangle - (-1)^{i+j} \langle (D_{i,A} - D_{i,B}) \rangle \langle (D_{j,A} - D_{j,B}) \rangle \nonumber \\
    &=& (-1)^{i+j} \sum_{\alpha} \Big[ \langle c_{i,\alpha}^{\dagger} c_{j+1,\alpha}^{} \rangle \langle c_{i+1,\alpha}^{} c_{j,\alpha}^{\dagger} \rangle + \langle c_{i,\alpha}^{\dagger} c_{j,\alpha}^{} \rangle \langle c_{i+1,\alpha}^{} c_{j+1,\alpha}^{\dagger} \rangle \nonumber \\
    && \qquad{} \qquad{} \qquad{} \qquad{} + \langle c_{i+1,\alpha}^{\dagger} c_{j+1,\alpha}^{} \rangle \langle c_{i,\alpha}^{} c_{j,\alpha}^{\dagger} \rangle + \langle c_{i+1,\alpha}^{\dagger} c_{j,\alpha}^{} \rangle \langle c_{i,\alpha}^{} c_{j+1,\alpha}^{\dagger} \rangle \Big] \nonumber \\
    &=& 4 (-1)^{r+1} \left( \frac{\sin^{2}(\pi r/2)}{(\pi r)^{2}} + \frac{\sin[\pi(r+1)/2]\sin[\pi(r-1)/2]}{\pi^{2}(r+1)(r-1)} \right) \,.
\end{eqnarray}

\subsection{Finite-bond and finite-size extrapolations of the correlation functions at the deconfined critical point}

\begin{figure*}[tb]
    \includegraphics[width=0.7\linewidth]{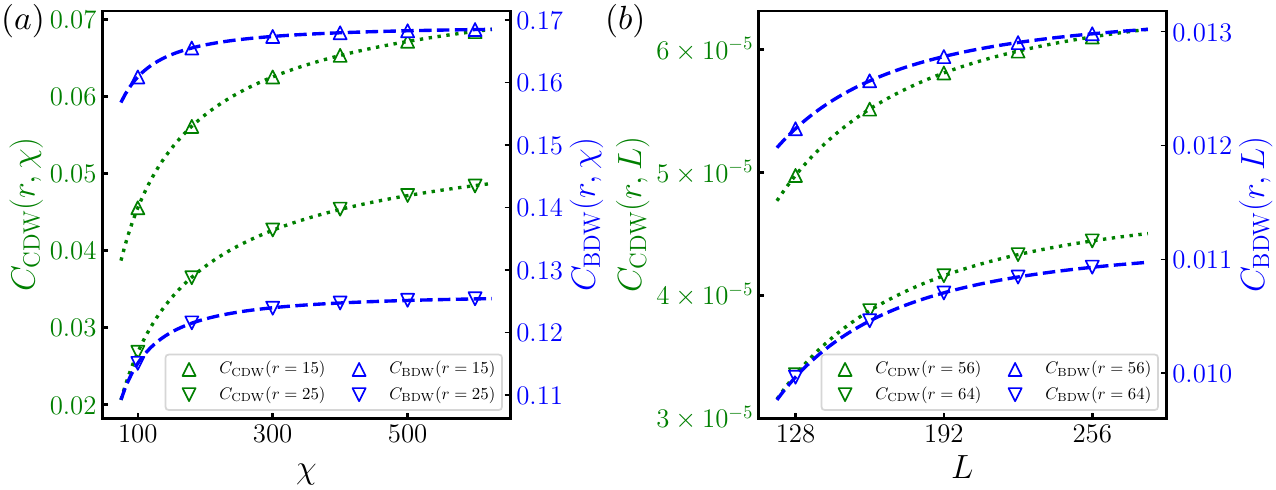}
    \caption{(a) Each data point of the bulk-bulk correlation function for BDW and CDW orders shown in Fig.2(b) in the main text is obtained by extrapolating to infinite MPS bond dimensions. The dotted and dashed lines are least-squares fittings with the form $C(r,\chi) = a (100/\chi)^{b} + C(r,\chi\to\infty)$. (b) Each data of the boundary-bulk correlations displayed in Fig.2(c) in the main text is obtained by extrapolating to $L\to\infty$ with the fitting form, $C(r,L) = a / L^{b} + C(r,L\to\infty)$, where $L$ is the ladder length. Numerical data are obtained from infinite-size DMRG calculations for (a) and finite-size DMRG calculations for (b) with a sufficiently large bond dimension $\chi=2048$ to ensure the convergence of the results for $L$ up to $256$.}
    \label{fig:extrapolation}
\end{figure*}

\begin{figure*}[tb]
    \includegraphics[width=1.0\linewidth]{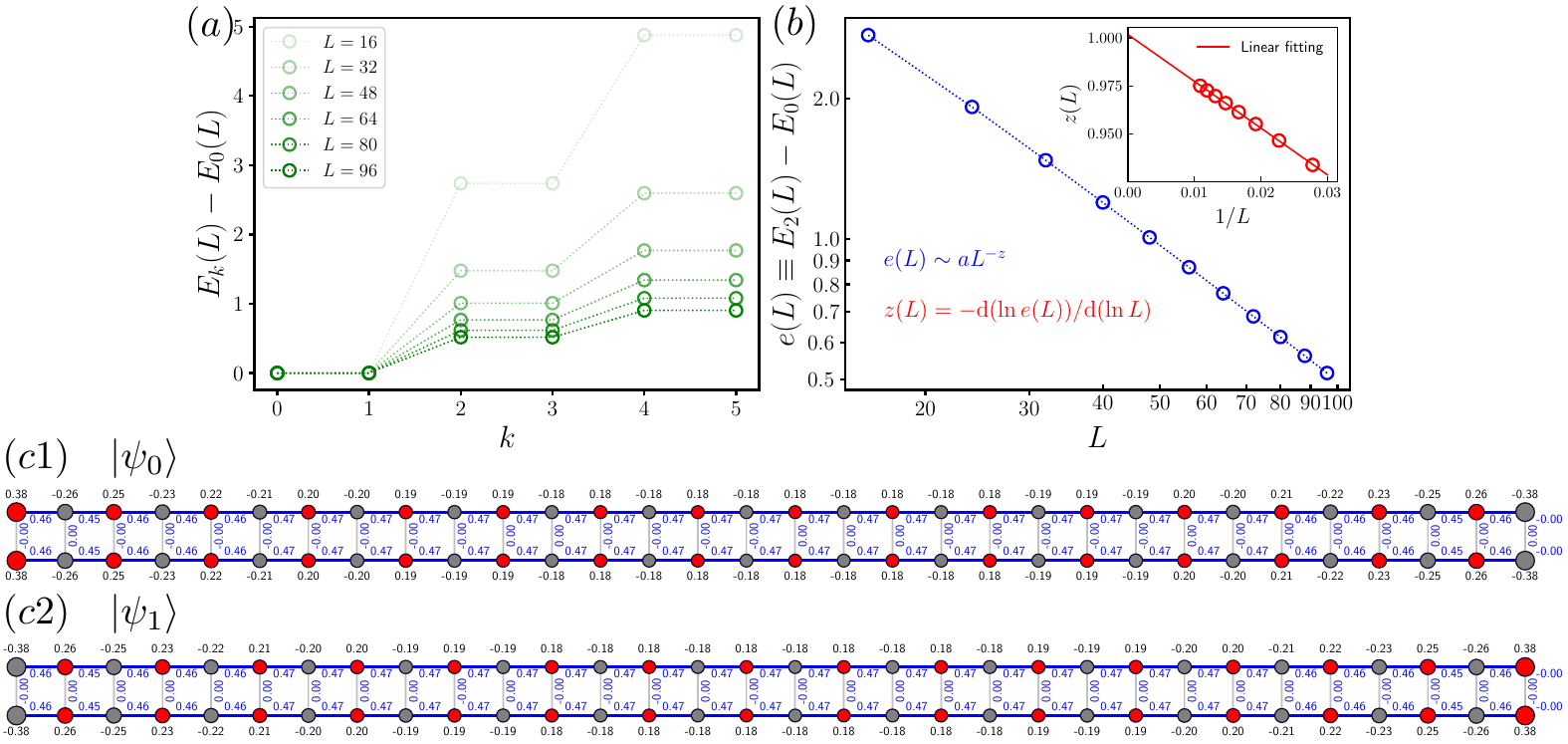}
    \caption{(a) The low-lying energy spectrum of the Hamiltonian at the estimated critical point $Q/t = 10$ and $V/t = 7.992$ under the open boundary condition for several system sizes. Only the first six energy levels are plotted here. (b) The finite-size scaling analysis for the energy gap $e(L) \equiv E_{2}(L) - E_{0}(L)$. Note the log-log scale used here. The inset shows the finite-size effect on the estimation of the dynamical critical exponent $z(L) = - \text{d}(\ln e(L))/\text{d}(\ln{L})$. A linear extrapolation gives $z(L\to\infty) = 1.002(1)$\,. (c1) and (c2) are intuitive illustrations of the double degenerate ground states for $L=32$. The values attached to the circles are $\langle{Z_{i,\alpha}}\rangle$ while the ones attached to the bonds are $\langle{(c^{\dagger}_{i,\alpha}c^{}_{j,\beta} + c^{\dagger}_{j,\beta}c^{}_{i,\alpha})}\rangle$ where $(i,\alpha)$ and $(j,\beta)$ are the sites connected by the bond. The double degenerate ground states are linearly combined in a proper way such that $|\psi_{0}\rangle$ and $|\psi_{1}\rangle$ are related by the bond-centered reflection $\mathbb{Z}_{2}^{B}$ symmetry.}
    \label{fig:gs_ill}
\end{figure*}

As shown in Fig.~2(b) in the main text, the bulk-bulk connected correlations are computed by infinite-size DMRG simulations. 
Although the thermodynamic limit is approached in a natural way, the calculation of the correlation functions is still affected by the finite-bond effect (or the finite-entanglement effect). 
To this end, we first perform a finite bond extrapolation for each $r$ of $C_\text{CDW/BDW}(r,\chi)$ to the limit $\chi \to \infty$ as shown in Fig.~\ref{fig:extrapolation}(a). 
The extrapolated results $C_\text{CDW/BDW}(r,\chi \to \infty)$ are then fed into the scaling analysis shown in Fig.~2(b) in the main text. 

However, the boundary-bulk correlations are obtained from finite-size DMRG calculations. 
Having chosen a sufficiently large MPS bond dimension, such as $\chi = 2048$, we can expect the results to converge well with $\chi$. 
In contrast to the bulk-bulk correlations, the accuracy of the boundary-bulk correlations are mainly limited by the finite system size. 
As shown in Fig.~\ref{fig:extrapolation}(b), the finite-size effect is eliminated here by extrapolating to the thermodynamic limit using $C_\text{CDW/BDW}(r, L)$ with $L$ up to $256$. 
Finally, the extrapolated results $C_\text{CDW/BDW}(r,L \to \infty)$ are used in Fig.~2(c) in the main text to study the boundary physics of the deconfined quantum critical point. 

\section{Existence of the fermion gapless edge modes at the deconfined critical point}

To reveal the existence of the fermion gapless edge modes at the deconfined critical point under investigation, we first calculate the low-lying energy spectrum of the model at the critical point estimated above under open boundary conditions. 
As shown in Figs.~\ref{fig:gs_ill}(a) and (b), it is clear that the ground state is double degenerate at $(V/t)_{c} = 7.992$ with $Q/t = 10$ and the energy gap defined by $e(L) \equiv E_{2}(L) - E_{0}(L)$ gradually closes as $L$ increases. 
A further finite-size scaling analysis finds that the gap scales as $e(L) \sim 1/L^{z}$ with a dynamical critical exponent $z = 1.002(1)$, which implies that the critical point is conformal. 
It is natural to expect that the double degeneracy of the ground state comes from the fermion gapless edge modes near the boundary, as all $\mathbb{Z}_{2}$ symmetries are restored at the critical point [both symmetry breaking orders vanish at the critical point; see Fig.~\ref{fig:idmrg_order_energy}(a)]. 
In Figs.~\ref{fig:gs_ill}(c) and (d), we further present the basic characteristics of the double degenerate ground states by calculating $\langle{Z_{i,\alpha}}\rangle$ for each site and $\langle{(c_{i,\alpha}^{\dagger}c_{j,\beta}^{} + c_{j,\beta}^{\dagger}c_{i,\alpha}^{})}\rangle$ for each bond. 
The CDW pattern that appears near the boundary is the result of the edge modes existing on the left or right boundary. 
It should be noted that the breaking of the symmetry near the boundary decays algebraically into the bulk [the boundary-bulk connected correlation $C_\text{CDW}(r)$ decays with a power law of $r$ as seen in Fig.~2(c) in the main text]; therefore, it does not contradict the restoration of symmetry in the bulk of the system. 

\begin{figure*}[tb]
    \includegraphics[width=1.0\linewidth]{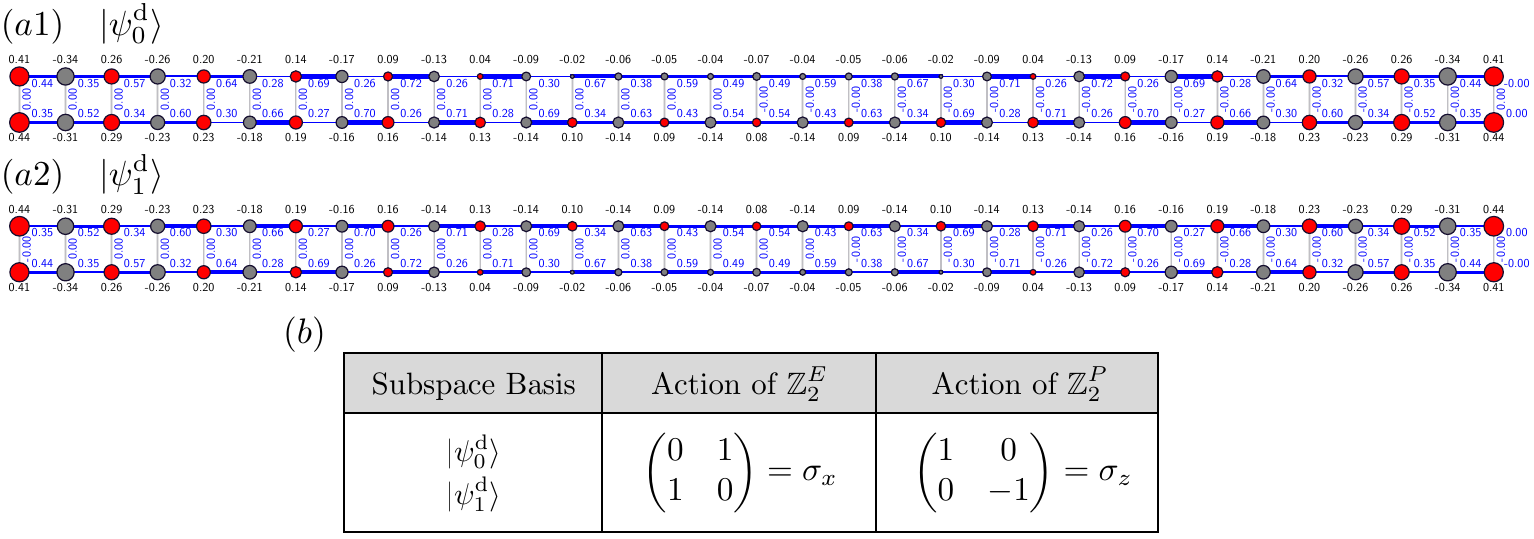}
    \caption{(a1) and (a2) give intuitive illustrations for the lowest-lying two degenerate eigenstates of the Hamiltonian with additional pinning fields, $-h \sum_{\alpha=A,B}(n_{1,\alpha} + n_{L,\alpha})$, applied on the boundaries. Simulations are performed for $L = 33$ and $h/t = 10$ under open boundary conditions. The meaning of the concrete numbers has been explained in the caption of Fig.~\ref{fig:gs_ill}. (b) summaries the resulting representation of the symmetry actions of layer-exchange $\mathbb{Z}_{2}^{E}$ and layer fermion parity $\mathbb{Z}_{2}^{P}$, within the two-dimensional subspace spanned by $\{ |\psi_{0}^\text{d}\rangle, |\psi_{1}^\text{d}\rangle \}$. The details of the symmetry analysis is given in Figs.~3(b) and (c) in the main text.}
    \label{fig:dw_ill}
\end{figure*}

On the other hand, the fermion gapless edge modes can be understood by introducing a CDW domain wall at the center of the two-leg ladder. 
More concretely, we modify the model Hamiltonian by applying additional pinning fields on the two boundaries as $H \rightarrow H - h \sum_{\alpha=A,B} (n_{1,\alpha} + n_{L,\alpha})$ with $h/t=10$.
By directly simulating the modified Hamiltonian at the estimated critical point $(V/t)_{c} = 7.992$ with $Q/t = 10$, we find that the degeneracy of the ground state is still twofold; Figs.~\ref{fig:dw_ill}(a1) and (a2) provide an intuitive illustration of the two states denoted by $|\psi_{0,1}^\text{d}\rangle$. 
To further study the interplay of various $\mathbb{Z}_{2}$ symmetries, we compute the effective representation of layer-exchange $\mathbb{Z}_{2}^{E}$ and layer fermi parity $\mathbb{Z}_{2}^{P}$ symmetries within the two-dimensional subspace spanned by $\{ |\psi_{0}^\text{d}\rangle, |\psi_{1}^\text{d}\rangle \}$. 
The details of the computation are shown in Figs.~3(b) and (c) in the main text. 
We find that both $\mathbb{Z}_{2}^{E}$ and $\mathbb{Z}_{2}^{P}$ act nontrivially within this subspace. 
More specifically, with a suitable linear combination of $|\psi_{0}^\text{d}\rangle$ and $|\psi_{1}^\text{d}\rangle$, the action of $\mathbb{Z}_{2}^{E}$ and $\mathbb{Z}_{2}^{P}$ can be represented exactly by Pauli matrices $\sigma_{x}$ and $\sigma_{z}$, respectively. 
As $\sigma_{x}$ and $\sigma_{z}$ anti-commute with each other, the action of $\mathbb{Z}_{2}^{E}$ and $\mathbb{Z}_{2}^{P}$ forms a projective representation of the $\mathbb{Z}_{2}^{E}\times \mathbb{Z}_{2}^{P}$. To sum up, the domain wall of $\mathbb{Z}_2^B$ broken phase carries the projective representation of $\mathbb{Z}_{2}^{E}\times \mathbb{Z}_{2}^{P}$, which is a manifestation of the mixed anomaly at the deconfined critical point.

\end{document}